# Particle-Hole Symmetry and the Fractional Quantum Hall Effect in the Lowest Landau Level

W. Pan[1,2,3], W. Kang[4], M.P. Lilly[3], J. L. Reno[3], K.W. Baldwin[5], K.W. West[5], L.N. Pfeiffer[5], and D.C. Tsui[5]

[1]Materials Physics Department, Sandia National Laboratories, Livermore, CA 94551 USA

[2]Quantum Phenomena Department, Sandia National Laboratories, Albuquerque, NM 87185 USA

[3]Center for Integrated Nanotechnologies, Sandia National Laboratories, Albuquerque, NM 87185 USA

[4]Department of Physics, University of Chicago, Chicago, IL 60637 USA

[5]Department of Electrical Engineering, Princeton University, Princeton, NJ 08544 USA

Abstract

We report on detailed experimental studies of a high-quality heterojunction insulated-gate field-effect transistor (HIGFET) to probe the particle-hole symmetry (PHS) of the FQHE states about half-filling in the lowest Landau level. The HIGFET was specially designed to vary the density of a two-dimensional electronic system under constant magnetic fields. We find in our constant magnetic field, variable density measurements that the sequence of FQHE states at filling factors $\nu = 1/3, 2/5, 3/7\ldots$ and its particle-hole conjugate states at filling factors $1 - \nu = 2/3, 3/5, 4/7\ldots$ have a very similar energy gap. Moreover, a reflection symmetry can be established in the magnetoconductivities between the $\nu$ and $1 - \nu$ states about half-filling. Our results demonstrate that the FQHE states in the lowest Landau level are manifestly particle-hole symmetric.

The fractional quantum Hall effect (FQHE) [1], found in a two-dimensional electron system (2DES) under low temperatures and high magnetic fields, is a unique, emergent state of matter with remarkable properties. As an incompressible quantum fluid of electrons, some of the exotic properties of the FQHE includes fractional charge [2], anyonic excitations [3,4,5], and non-Abelian statistics [6,7]. There has been a considerable interest in the realization of fault tolerant topological quantum computing to take advantage of the anyonic excitation in the FQHE [8,9].

Historically the most successful approach towards understanding the FQHE and related phenomena was based on the theory of Halperin, Lee, and Read (HLR) [10], which maps the electrons at half-filled Landau level to a system of composite fermions coupled to a Chern-Simons gauge field [11,12]. Such a singular gauge transformation serves to attach two flux quanta of magnetic field on the average to every electron, leading to a system of composite fermions that moves under a zero effective magnetic field at half-filling. The FQHE states at filling factors $\nu = p/(2p \pm 1)$, where $p = 1, 2, 3…$, can be understood as $\nu^* = p$ integer quantum Hall effect states of composite fermions. Here, $\nu = nh/eB$. $n$ is the 2DES density, $h$ the Planck constant, $e$ the electron charge, and $B$ the magnetic field.

Over the years experiments have provided a strong support for the existence of composite fermions at half-filling [13-16]. In turn, the composite fermion theory has been a very successful as a consistent framework to understand the experiments on the FQHE. Recently particle-hole symmetry (PHS) in the lowest Landau level [17] has generated a renewed interest in the physics of composite fermions at half-filling [18-30]. The HLR theory has largely neglected the presence of particle-hole symmetry even though a two-dimensional electronic system with a partially-filled

Landau level of spin-polarized electrons possesses an exact PHS about half-filling in the limit of zero effective mass.

In an approach that is distinct from the HLR theory, a new theory has conjectured that the composite fermions (CFs) at half-filled Landau level may be a Dirac particle [21]. As an effective theory of composite fermions at half-filling, the Dirac composite fermion theory shares a number of predictions about the experiments with the HLR theory. Since a Dirac particle possesses an inherent PHS about half-filling, the PHS in a Dirac composite fermions theory is explicit. Support for the Dirac composite fermion theory has come from studies of numerics. However, there is a general lack of understanding on how the Dirac theory of composite fermions may emerge from the microscopics starting from a Hamiltonian for two-dimensional electrons. It also remains unclear how the Dirac theory of composite fermions may be distinguished experimentally from the HLR theory.

In light of these developments in the theory of CFs, experimental clarification of PHS of FQHE is necessary. Although the PHS between the $\nu$ and 1- $\nu$ FQHE states is often taken for granted, a *direct* experimental confirmation of PH symmetry has not been made to date. Since most modulation doped GaAs/AlGaAs heterostructures have a constant density of carriers, experimental study of the $\nu$ and 1- $\nu$ FQHE states in a constant density specimen necessitates measurements at different magnetic fields. Previous experiments on constant density systems have only indirectly probed the PHS between the $\nu$ and 1- $\nu$ FQHE states in the lowest Landau level [31,32]. Their interpretations are also complicated by the energy gaps being modified by variations in the wave function, magnetic length, the Landau level mixing and the effect of disorder potential as magnetic field is varied.

We have carried out detailed experimental studies of electronic transport properties in specially designed heterojunction insulated-gate field-effect transistor (HIGFET) [33] in GaAs/AlGaAs heterostructures to probe the PHS between the FQHE states about the half-filling. In our experiment we are able to measure and compare the energy gaps ($E_g$) of the FQHE states at ν and 1 - ν at the same magnetic field, allowing for a direct comparison of the PH-symmetric FQHE states in the lowest Landau level.

Fig. 1a shows schematically the device structure of a HIGFET used in this study. A heavily n-doped GaAs gate layer above a molecular-beam epitaxy (MBE)-grown AlGaAs insulating layer enables a realization of uniform, high mobility 2DES as a function of gate voltage. Fig. 1b shows the magnetoresistance, $\rho_{xx}$, traces as a function of electron density $n$ in HIGFET A, measured at a fixed magnetic field of 6T and at four separate temperatures. Well-developed FQHE states are found at the fillings ν = 1/3, 2/5, 3/7 and 4/9 for ν < 1/2 and ν = 2/3, 3/5, 4/7 and 5/9 for ν > 1/2. The energy gap, $E_g$, of these FQHE states are obtained from the exponential dependence of $\rho_{xx}$ as a function of temperature: $\rho_{xx} \sim \exp(-E_g/k_B T)$. Arrhenius plots of three strongest FQHE states above and below ν = 1/2 are shown in Fig. 1c. Semilog plots of $\rho_{xx}$ as a function of 1/$T$ for each of the FQHE states at fillings ν and 1- ν overlap almost perfectly. This demonstrates that the conjugate pair states at filling states ν and 1- ν have essentially the same energy gap, as required by PHS. Their overlaps also imply that the disorder broadening for the ν and 1 - ν states is nearly the same and particle-hole symmetric. From the slope of the linear fit, the energy gaps of the FQHE states can be obtained.

We have undertaken extensive temperature-dependent measurements of $\rho_{xx}$ at other magnetic fields, data at $B = 4.5$, 9, and 13.5T shown in Fig. S1 in the supplementary materials. In all cases, we find that the pairs of FQHE states at fillings ν and 1 - ν states have very similar energy gaps. The variation is within 10% or less, as long as the magnetic field is the same.

To further elucidate our energy gap measurements, all the energy gap values of the FQHE states in the lowest Landau level can be compared quantitatively with theoretical predictions. The energy gaps for the pair of ν and 1- ν FQHE states are predicted to be [10]:

$$E_g(v) = \frac{g^*}{|2p+1|} \frac{e^2}{\varepsilon l_B} \tag{1}$$

where the constant $g^* = g^v$ for ν < 1/2 and $g^* = g^{1-v}$ for ν > 1/2, $p = \pm 1, \pm 2, \pm 3\ldots$, $\varepsilon = 12.8$ is the dielectric constant of GaAs, and $l_B = (\hbar/eB)^{1/2}$ is the magnetic length. For ν < 1/2, $p$ is a positive integer; for ν > 1/2, $p$ a negative integer. PHS requires that $g^v = g^{1-v}$ and the energy gaps of the ν and 1- ν states must be equal under a constant magnetic field.

Fig. 2 shows the measured energy gap $E_g$ vs $B^{1/2}/(2p+1)$, illustrating a universal scaling of the two sets of energy gap values on either side of ν = 1/2 into a set of symmetric lines about the vertical with nearly equal but opposite slope. A high level of reflection symmetry about $B^{1/2}/(2p+1)$ = 0 is a direct consequence of PHS about ν = 1/2 [34].

We find that empirically $g^* \approx 0.15$ for both left and right sides, approximately one half of the predicted value [10]. The reduction in the value of $g^*$ relative to the predicted value is generally attributed primarily to the effect of finite thickness of the 2DES [35,36,37].

Because of disorder broadening of the energy gap of the FQHE states, we find that Eq. (1) is modified by a negative intercept $\Gamma_v$ with $E_g(v) = (\frac{g^*}{|2p+1|})(e^2/\varepsilon l_B) - \Gamma_v$ and $\Gamma_v \sim 1.8$K. We also find that the effect of disorder is particle-hole symmetric with $\Gamma_v = \Gamma_{1-v}$. Existence of the disorder broadening in the energy gap energy has been observed in the past studies of the FQHE energy gap [32,38,39]. Similar disorder broadening is observed in spite of very different mobilities and the type of carriers. This seems to suggest that this disorder broadening is independent of the residual disorder experienced by the underlying electrons or holes. This is a perplexing phenomenon that still needs an explanation [40].

We note a slight difference of the energy gaps for the $\nu = 1/3$ and $2/3$ FQHE states may be attributed to the skyrmions observed in the 1/3 FQHE state [41] and merits a further investigation. Since the energy gaps of the FQHE states can be interpreted as the effective cyclotron energy of composite fermions [10], the same slopes above and below $\nu = 1/2$ also indicates that the effective mass is the same for both the particle and hole states of CFs in the lowest Landau level. Indeed, the effective cyclotron energy of CFs can be written as $\hbar e B_{eff}/m^*$ [10] where $B_{eff}$ is the effective magnetic field of composite fermions and $m^*$ is the effective mass of composite fermions with $m^* \propto B^{1/2}$ [10]. Under a constant magnetic field, $B_{eff} = B/(2p+1)$ at fillings of $\nu = p/(2p+1)$. It follows that the cyclotron energy of CFs $\propto B^{1/2}/(2p+1)$. From the values of the $g^*$, we obtain an effective mass of $m^* \approx 0.2\, m_e$, which is consistent with the effective mass obtained in previous studies in fixed density 2DES samples [38,39].

One physical parameter that may affect the energy gap of the FQHE states is the Landau level mixing, which mixes states in the lowest Landau level with those from higher lying Landau levels. The degree of Landau level mixing, $\kappa$, is parametrized by the ratio of Coulomb energy (which is proportional to $B^{1/2}$) over Landau level separation (which is proportional $B$). For example, $\kappa = 1.2$ at $B = 4.5T$ and $\kappa = 0.7$ at $13.5T$. The scaling of the energy gap data at various magnetic fields in Fig. 2 suggests that the Landau level mixing effect is expected to play a minor role in affecting the excitation gaps at ν and 1-ν [42].

In addition to the energy gap data, another evidence of the PHS of the FQHE comes from an unexpected reflection symmetry of magnetotransport coefficients between the regions of ν < 1/2 and ν > 1/2. In standard transport theories, such as in Boltzmann's linear response theory or in the Kubo formalism, conductivities are the direct outcomes of a diffusive transport [43]. In order to reveal PHS, one needs to convert magneto-resistivity and Hall resistivity (experimentally measured, shown in Fig. S2) to magneto-conductivity $\sigma_{xx}$ and Hall conductivity $\sigma_{xy}$, using the formula $\sigma_{xx} = \rho_{xx}/(\rho_{xx}^2+\rho_{xy}^2)$ and $\sigma_{xy} = \rho_{xy}/(\rho_{xx}^2+\rho_{xy}^2)$.

Fig. 3a and 3b show, respectively, $\sigma_{xx}$ and $\sigma_{xy}$ in another HIGFET, as a function of Landau level filling at a fixed $B$ field of 6T. Examining the $\sigma_{xx}(\nu)$ trace carefully, the whole trace is remarkably symmetric around ν = 1/2. In fact, the reflection symmetry is clearer about the 1/2 state if the $\sigma_{xx}$ trace is reversed and overlaid on top of itself. As shown in Fig. 3c, $\sigma_{xx}$ above ν > 1/2 overlaps almost perfectly with $\sigma_{xx}$ above ν < 1/2 and vice visa. This overlap explicitly demonstrates PHS between the particle-hole conjugate ν and 1- ν FQHE states. Having revealed the unexpected reflection symmetry in $\sigma_{xx}$, we now turn to $\sigma_{xy}$. In Fig. 3d traces of $\sigma_{xy}(\nu)$ and 1 - $\sigma_{xy}(1-\nu)$ are compared in Fig. 3d. The two curves overlap very well, particularly in the gapped FQHE states.

However, the overlap is not perfect with slight discrepancy between some $\sigma_{xx}$ peaks such as those found between ν = 1/3 and 2/5 and between 2/3 and 3/5. The minimum between the 2/3 and 3/5 states does not occur at its PH conjugate position between 1/3 and 2/5. This minimum is probably due to a developing ν =7/11 state, which has also been observed in single junction samples of similar electron mobility before [44]. Its particle-hole conjugate state, the 4/11 state [45,46], is known to be very fragile and only exists in ultra-high mobility wide quantum well samples. It follows that the smallness of the energy gaps for the ν = 4/11 and 7/11 states makes it difficult to compare fully formed states in our sample.

Fig. 4 shows an evolution of the reflection symmetry in $\sigma_{xx}$ over a range of magnetic fields from 1.5T to 12T. $\sigma_{xx}$ in lower magnetic fields are complicated by a stronger Landau level mixing effect as well as spin transitions. Under 1.5T (where κ = 2.1) a large difference in $\sigma_{xx}$ can be found between ν and 1- ν states. Reflection symmetry is fully realized at larger magnetic field where the energy gaps of the FQHE states are larger and the effect of Landau level mixing may be negligibly small. We note that type of reflection symmetry cannot be observed in magnetic field sweeps in constant density sample.

In conclusion, we have explored the particle hole symmetry of the FQHE in the lowest Landau level of a 2DES. We have observed nearly identical energy gaps of the FQHE states at fillings ν = 1/3, 2/5, 3/7… and their particle-hole conjugate states at 1 - ν = 2/3, 3/5, 4/7… We also observe a striking reflection symmetry in the magneto-conductivities between the ν and 1- ν states about half-filling. Our results demonstrate that particle-hole symmetry is a fundamental property of the FQHE states in the lowest Landau level.

We thank K. Yang and P.A. Sharma for helpful discussions. This work was supported by a Laboratory Directed Research and Development (LDRD) project at Sandia National Laboratories. Materials growth and device fabrication at Sandia was supported by U.S. Department of Energy (DOE), Office of Science, Basic Energy Sciences, Materials Sciences and Engineering Division and performed at the Center for Integrated Nanotechnologies, an Office of Science User Facility operated for the DOE Office of Science. Sandia National Laboratories is a multimission laboratory managed and operated by National Technology and Engineering Solutions of Sandia, LLC., a wholly owned subsidiary of Honeywell International, Inc., for the U.S. Department of Energy's National Nuclear Security Administration under contract DE-NA-0003525. This paper describes objective technical results and analysis. Any subjective views or opinions that might be expressed in the paper do not necessarily represent the views of the U.S. Department of Energy or the United States Government. Part of our measurements were performed at the National High Magnetic Field Laboratory (NHMFL), which is supported by the NSF Cooperative Agreement DMR 1157490, by the State of Florida, and the DOE. The work at University of Chicago was supported in part by the Templeton Foundation and NSF MRSEC Program through the University of Chicago Materials Center. Sample growth and device fabrication at Princeton was funded by the Gordon and Betty Moore Foundation through the EPiQS initiative GBMF4420, and by the National Science Foundation MRSEC Grant DMR-1420541.

**Figures and captions**

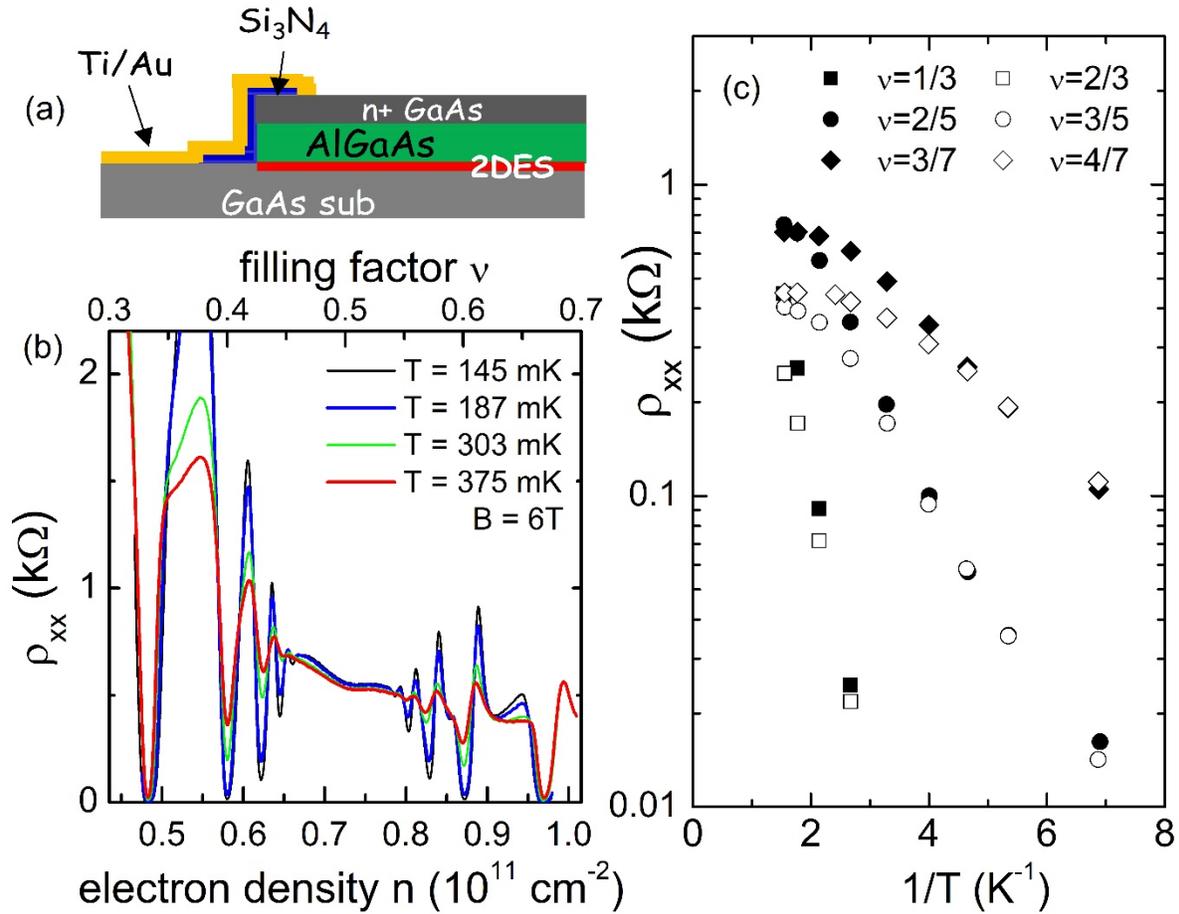

Figure 1. (a): Schematic of the MBE device structure of a HIGFET used in this study is shown. A heavily n-doped GaAs gate layer above a molecular-beam epitaxy (MBE)-grown AlGaAs insulating layer enables a realization of uniform, high mobility 2DES as a function of gate voltage. (b): Magnetoresistance, $\rho_{xx}$, trace as a function of electron density in HIGFET A, measured at a fixed magnetic field of 6T. The corresponding values of Landau level filling factors are shown at the top. (c): Arrhenius plots of the three strongest FQHE states above and below $\nu = 1/2$.

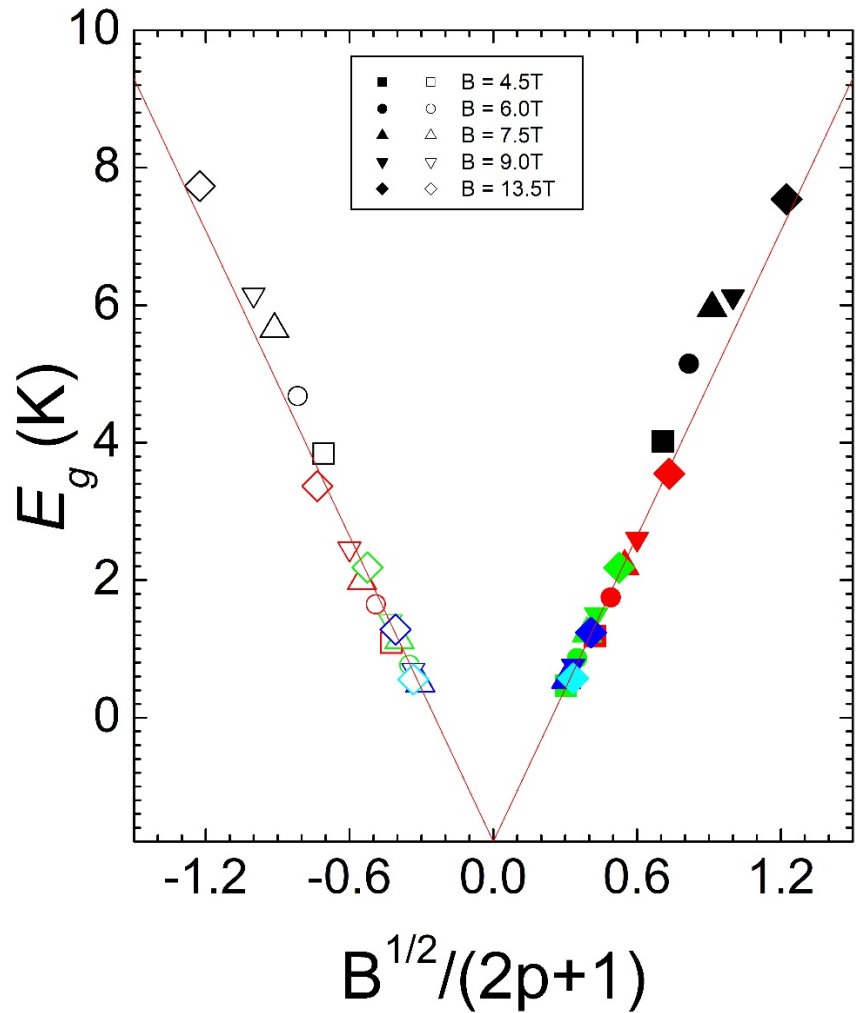

Figure 2. Energy gaps $E_g$ of FQHE states vs $B^{1/2}/(2p+1)$ measured at various magnetic fields. The solid symbols are for $\nu <1/2$ and open symbols for $\nu > 1/2$. The black color represents the third states (1/3 and 2/3), red the fifth states (2/5 and 3/5), green the seventh states (3/7 and 4/7), blue the ninth states (4/9 and 5/9), and cyan the eleventh states (5/11 and 6/11). The universal scaling of the energy gap values on either side of $\nu = 1/2$ is indicated by a set of symmetric lines (red color) about the vertical with nearly equal but opposite slope.

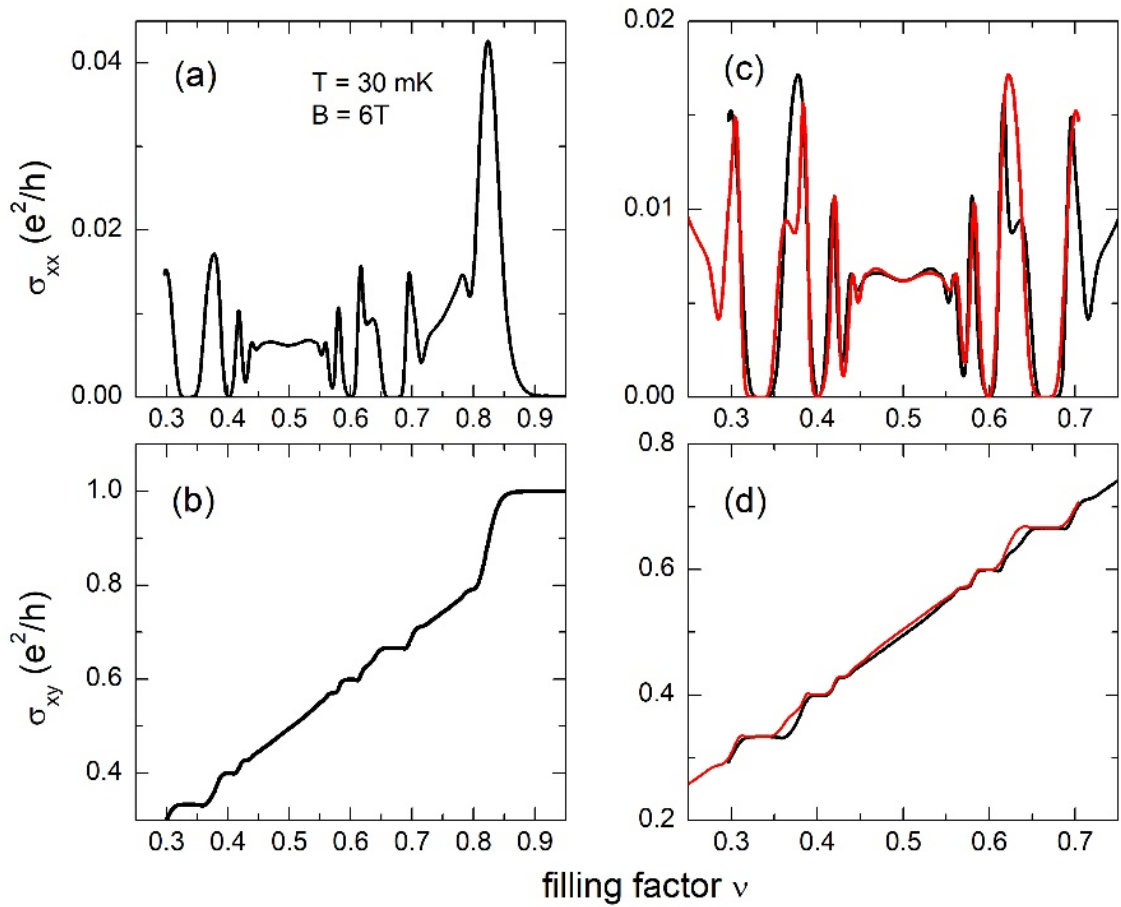

Figure 3. (a): Longitudinal conductivity $\sigma_{xx}$, converted from the measured resistivities $\rho_{xx}$ and $\rho_{xy}$, around $\nu = 1/2$ at $B = 6T$. (b): Hall conductivity $\sigma_{xy}$, converted from the measured resistivities $\rho_{xx}$ and $\rho_{xy}$, around $\nu = 1/2$ at $B = 6T$. (c): reflection symmetry in $\sigma_{xx}$ about the 1/2 state where the $\sigma_{xx}$ trace is reversed (red curve) and overlaid on top of itself (black curve). (d): reflection symmetry in $\sigma_{xy}$ about the 1/2 state where the $\sigma_{xy}(\nu)$ trace (black curve) and the 1 - $\sigma_{xy}(1-\nu)$ trace (red curve) overlap each other.

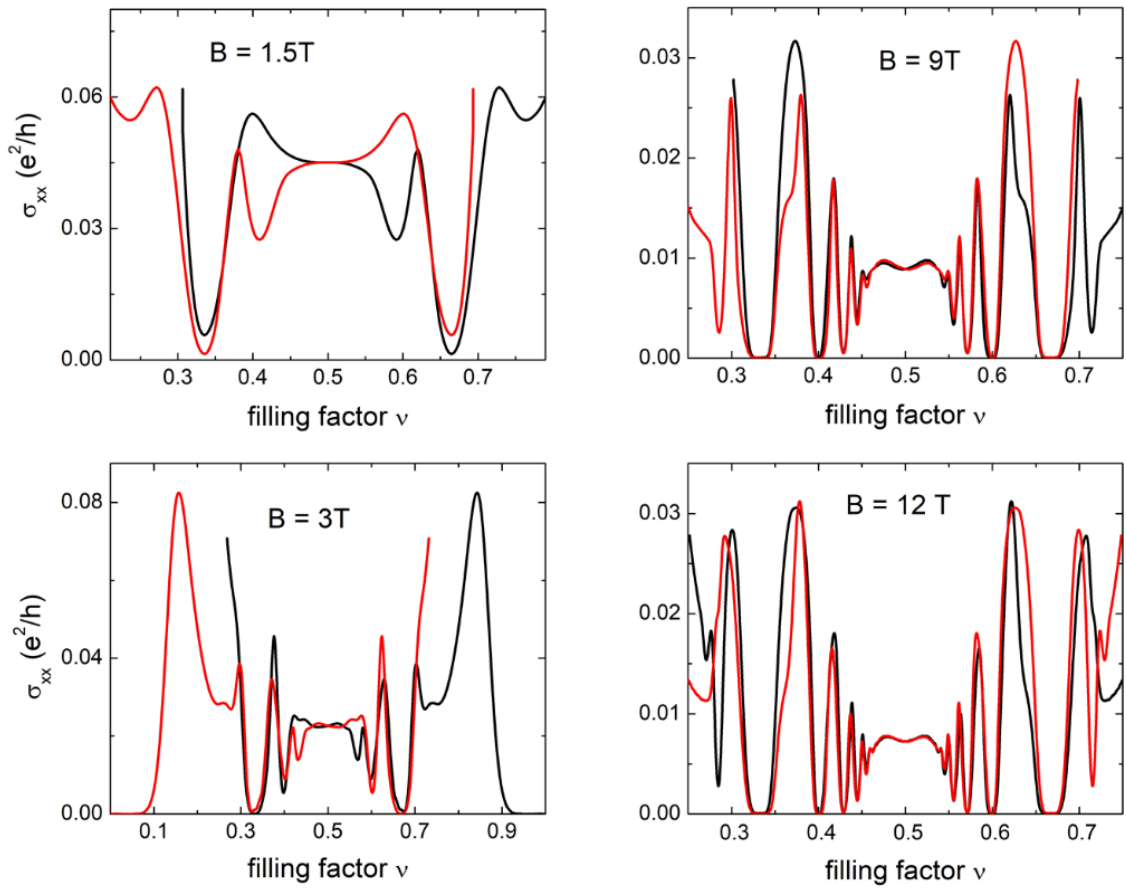

Figure 4. Examination of reflection symmetry in $\sigma_{xx}$ over a large range of magnetic fields, from 1.5T to 12T.

**Supplementary Materials**

The heterojunction insulated-gated field-effect transistors (HIGFETs) [33] were exploited for the quantum Hall measurements. Low frequency (~ 11 Hz) lock-in (Princeton Applied Research 124A) technique was used to collect $R_{xx}$ and $R_{xy}$ as a function of electron density by sweeping the gate voltage at a fixed magnetic (B) field. $\rho_{xx}$ is obtained from the measured $R_{xx}$ by taken into account the geometric ratio, and $\rho_{xy} = R_{xy}$ in two-dimensions.

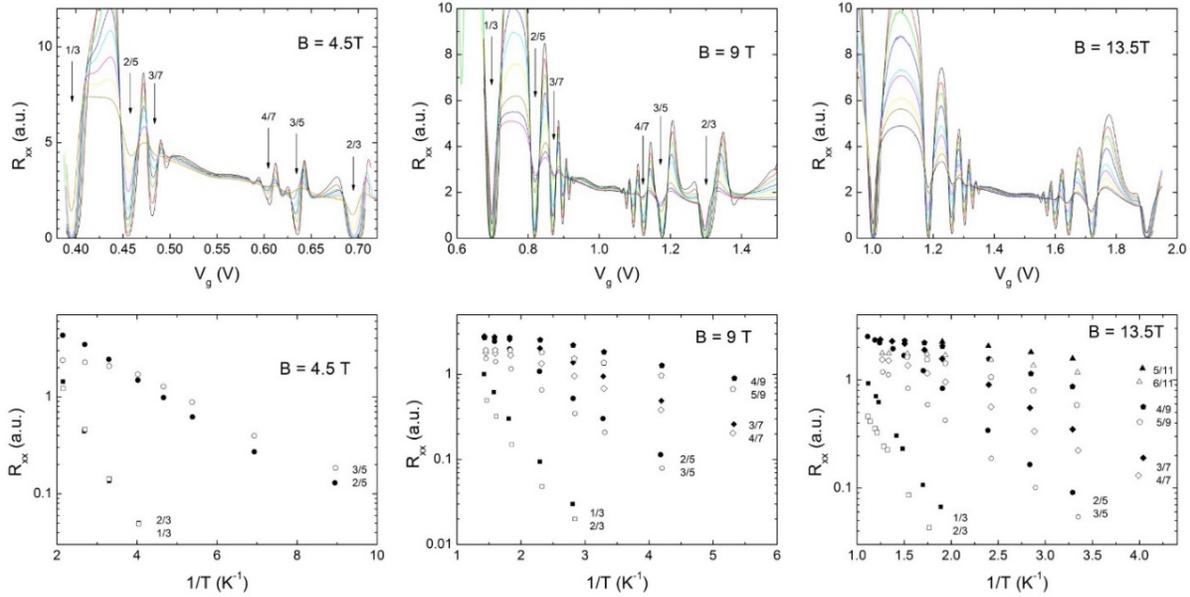

Figure S1

Fig. S1 show the temperature dependence of Rxx and the activation plots are various magnetic field. Again, at each magnetic field, the particle-hole conjugate states display the same T dependence. In other words, their energy gaps are the same.

Fig. S2 show the magneto-resistivity $\rho_{xx}$ and Hall resistivity $\rho_{xy}$ data used to construct Fig. 3. Representative FQHE states are marked by arrows. In these traces, the PH symmetric features are not properly revealed. For example, the value of the resistivity peak between the nu=2/5 and 3/7 FQHE states is much higher than that between 3/5 and 4/7. This is not surprising. As we point out in the main text, in standard transport theories, such as Boltzmann's linear response theory or the Kubo formalism,

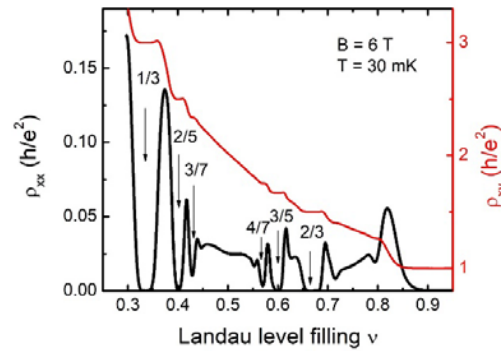

Figure S2

conductivities are the direct outcomes of a diffusive transport [43].

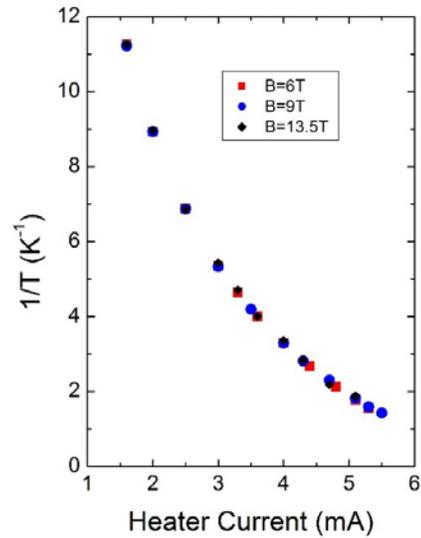

In our measurements, temperature was measured by a calibrated $RuO_2$ thermometer fixated very close to the samples. The calibration was done against a $^3$He melting curve thermometer and is very accurate at zero magnetic field. However, its resistance varies with the magnetic field. Thus, magnetoresistance correction to the thermometer is important in obtaining an accurate reading of fridge temperature and, consequently, in obtaining accurate FQHE energy gaps. To this end, we first set the current to the heater nearby the thermometer and waited for at least 1.5 hours until the fridge temperature was stable. We then swept the magnetic field slowly to 13.5T. In so doing, we obtained a matrix of RuO2 resistance versus heater current and magnetic field [S1]. Based on this matrix, we can make an accurate intercalation for any heater current and magnetic field. Fig.S3 shows the reversed temperature as a function of heater current at three selected magnetic fields. All the data points collapse onto a single curve, thus demonstrating that the magneto-resistance correction to the thermometer has properly been taken care of.

Figure S3